\documentclass[prl,twocolumn,showpacs,fleqn]{revtex4-1}

\usepackage{graphicx}
\usepackage{amsmath}
\usepackage{amsfonts}
\usepackage{amssymb}
\usepackage[svgnames]{xcolor}
\newcommand{\braket}[2]{\left\langle #1|#2 \right\rangle}

\newcommand{\mean}[1]{\left\langle #1 \right\rangle}
\begin{document}
\title{Nonlinear phononics using atomically thin membranes}
\author{Daniel Midtvedt}
\email{midtvedt@pks.mpg.de}
\author{Andreas Isacsson}
\author{Alexander Croy}
\email{croy@pks.mpg.de}

\affiliation{Department of Applied Physics,
Chalmers University of Technology, S-41296 G\"{o}teborg, Sweden}

\date{\today}

\pacs{%
}

\begin{abstract}\noindent
Phononic crystals and acoustic meta-materials are used to tailor phonon and sound propagation properties by utilizing artificial, periodic structures. Analogous to photonic crystals, phononic band gaps can be created, which influence wave propagation and, more generally, allow engineering of the acoustic properties of a system. Beyond that, nonlinear phenomena in periodic structures have been extensively studied in photonic crystals and atomic Bose-Einstein Condensates in optical lattices.
However, creating nonlinear phononic crystals or nonlinear acoustic meta-materials remains challenging and only few examples have been demonstrated. Here we show that atomically thin and periodically pinned membranes support coupled localized modes with nonlinear dynamics. The proposed system provides a platform for investigating nonlinear phononics.
\end{abstract}

\maketitle

%%%%%%%%%%%%%%%%%%%%%%%%%%%%%%%%%%%%%%%%%%%%%%%%%%%%%%%%%%%%%%%%%%%%%%%%%%%%%%%
%% INTRO
%%%%%%%%%%%%%%%%%%%%%%%%%%%%%%%%%%%%%%%%%%%%%%%%%%%%%%%%%%%%%%%%%%%%%%%%%%%%%%%
Nonlinear photonic\cite{sojo04} and optomechanical crystals\cite{eich+09,helu+11,luma+13}, as well as atomic Bose-Einstein Condensates (BEC) in optical lattices\cite{jabr+98,bl05,moob06} are well-known examples for which nonlinear phenomena in periodic structures have been intensively studied\cite{beiz+05}. 
All these systems exploit the wide tunability of system properties offered by artificial periodic structures. Similarly, phononic crystals can be engineered to have tailored phonon propagation properties\cite{lufe+09,bani12,ma13}. However, nonlinear effects in phononic crystals are, in general, much harder to obtain\cite{both+11}.

The basic building block of a phononic crystal consists of two coupled localized phonon modes, a phononic dimer. From studies of BECs\cite{smfa+97,alga05,zini+10}
and nanomechanical resonators\cite{shim+07,licr08,kacr+09,tsmo+13} it is known, that the presence of nonlinearities leads to a rich dynamical behavior even for such a simple system. Moreover, coherent manipulation of phonons in coupled resonators has recently been demonstrated
in the linear ``Rabi'' regime\cite{okgo+13,fari+13}. 
Accessing the nonlinear regime and the ability to scale the system beyond dimers will be crucial for realizing nonlinear phononics. Nanomechanical systems which exhibit a combination of strong nonlinearities and a high level of control and tunability\cite{shim+07,licr08,kacr+09,tsmo+13,okgo+13,fari+13,fari+12,lewe+08,limi+12,sook+12,eimo+11,wegu+14}, are promising candidates in this respect.

Here we show that atomically thin and periodically pinned membranes provide an excellent platform for nonlinear phononics. In particular, they offer high scalability and facilitate the design of arrays of defects. Using a continuum mechanics description \cite{atis+08} of a pinned graphene membrane we find that vacancies in the pinning lattice support localized flexural modes which can be accessed and tuned individually. Two such modes in close proximity interact via the elastic energy and constitute a simple phononic dimer. We show that the system can be tuned from the linear ``Rabi'' to the ``Josephson'' regime\cite{smfa+97,zini+10}. Based on the dimer results, our setup can be extended to construct arrays of defects, which support localized modes with adjustable properties. The nonlinear interaction between the localized flexural modes provides a viable path to study phononic many-body effects. Due to the high frequencies of the modes and the low mass of the membrane, our system is an ideal candidate for studies in the quantum regime\cite{poza12}.

%%%%%%%%%%%%%%%%%%%%%%%%%%%%%%%%%%%%%%%%%%%%%%%%%%%%%%%%%%%%%%%%%%%%%%%%%%%%%%%
%% MODEL
%%%%%%%%%%%%%%%%%%%%%%%%%%%%%%%%%%%%%%%%%%%%%%%%%%%%%%%%%%%%%%%%%%%%%%%%%%%%%%%
\begin{figure}[b]
	\centering
	\includegraphics[width=\columnwidth]{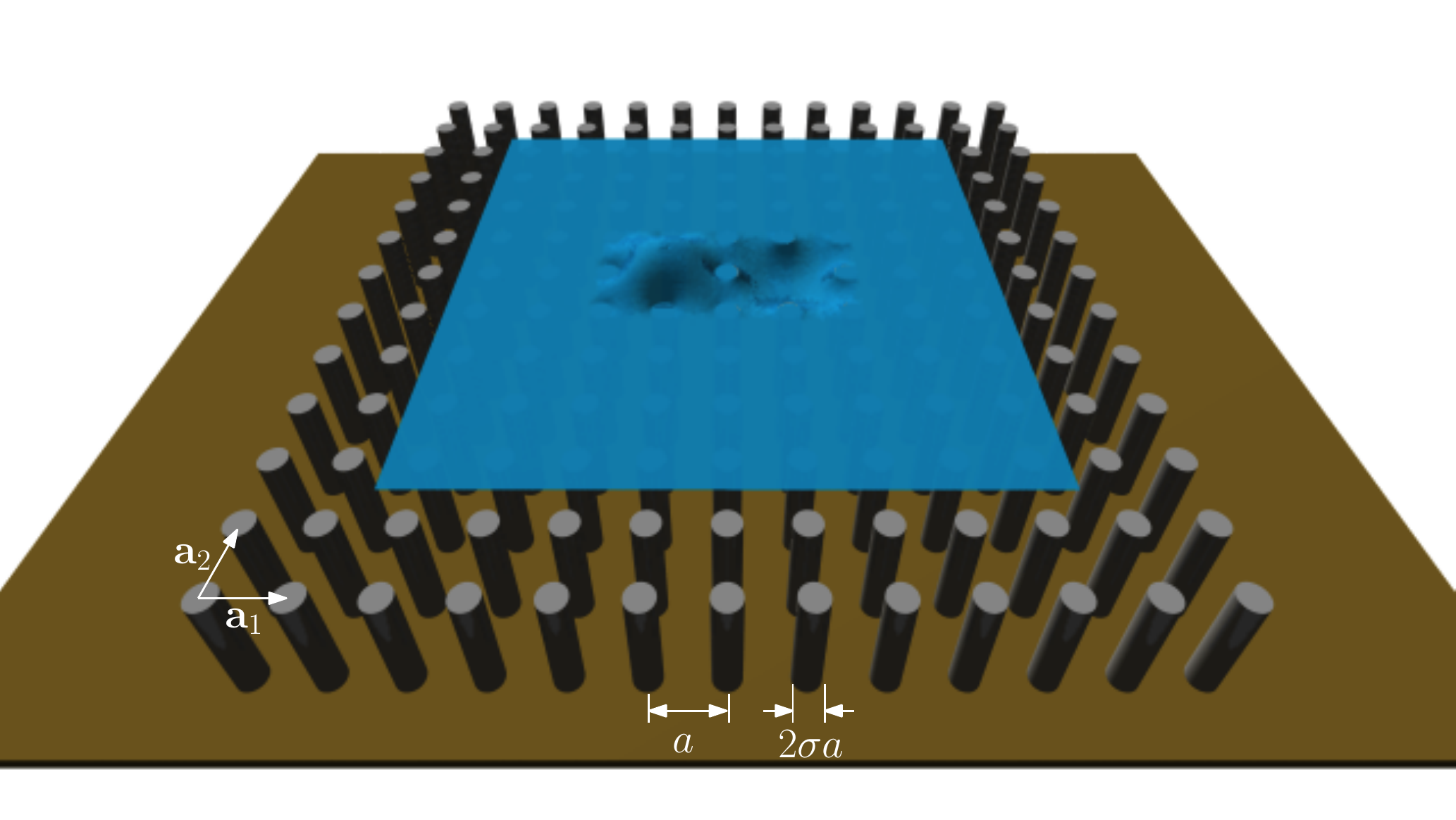}
    \caption{Schematic illustration of the pinning-lattice setup: Cylindrical pillars with diameter $2\sigma a$ are arranged
        in a periodic lattice with lattice spacing $a$. The suspended graphene membrane is pinned at the top of the pillars.
        By removing pillars, localized modes can be created.}
    \label{fig:setup}
\end{figure}%
%%%%%%%%%%%%%
\section{Results}
\noindent
{\bf Pinned graphene membrane.} In order to realize the phononic crystal, we consider a graphene membrane, which is deposited on top of
a square lattice of cylindrical pillars with radii $\sigma a$ and lattice vectors $\mathbf{a}_1$ and $\mathbf{a}_2$ (see Fig.\ \ref{fig:setup}). Due to the large adhesion energy,
the graphene membrane is essentially pinned at the pillars\cite{kuca+11,mebl13}, but free to move otherwise. This
situation may be described using a continuum mechanics model\cite{atis+08,kuca+11}. The Lagrangian density
of a graphene sheet subject to a pinning potential $V(\mathbf{r})$ is
\begin{equation}\label{eq:Lagrangian}
    \mathcal{L}=\frac{1}{2} \rho \dot{w}^2 - \frac{\kappa}{2} (\nabla^2 w)^2 - \frac{1}{2}\sigma_{ij}\epsilon_{ij} -\frac{1}{2} V(\mathbf{r}) w^2	\quad(i,j=x,y)\;,
\end{equation}
where $w(\mathbf{r})$ is the out-of-plane deformation, $\rho$ is the mass density, $\kappa$ is the bending rigidity,
$\sigma_{ij}$ is the stress and $\epsilon_{ij}$ is the strain tensor\cite{lali86}. 
Here we have used the out-of-plane approximation\cite{atis+08},
which is valid for large pre-strain typically found for suspended graphene sheets\cite{rofa+11}. For isotropic pre-strain, $\epsilon_{ij}\to\varepsilon \delta_{i,j} + \partial_i w \partial_j w/2$, the Lagrangian density can be further simplified. The pre-tension
$T_{0}=\varepsilon(\lambda + 2\mu)$ determines the unit of frequency and $\lambda$ and $\mu$ are the Lam\'e parameters of the membrane. The pinning potential is given by $V(\mathbf{r})=V_0 \sum _{nm} \Theta(|\mathbf{r}-\mathbf{r}_{nm}|-\sigma a)$, where $V_0$ is the pinning strength and $\mathbf{r}_{nm}=n\cdot\mathbf{a}_1+ m \cdot \mathbf{a}_2$. Due to the periodicity of the pinning lattice, the system supports frequency bands analogous to the electronic bands in periodic solids. See Supplementary Note 1 for a calculation of the lowest lying frequency bands in the present system.

In the following we take the lattice spacing $a$ to be the unit of length, $a \sqrt{\rho/T_0}$ the unit of time and
$\rho a^2$ the unit of mass.\\

\begin{figure*}[ht]
	\centering
	\includegraphics[width=\textwidth]{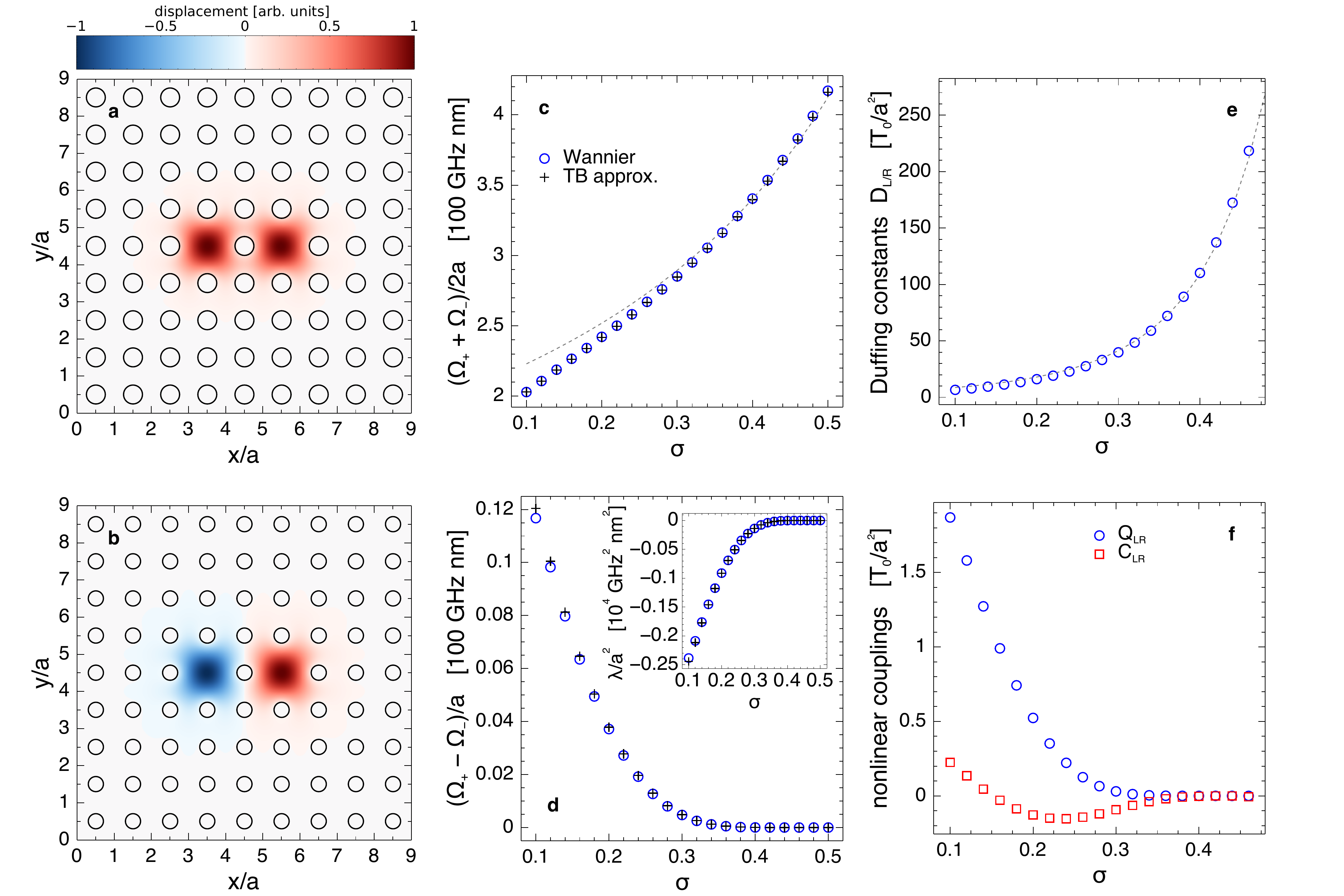}
	\caption{Coupled localized modes: Displacement field of (a) symmetric and (b) anti-symmetric modes in a lattice with two vacancies. 
			The super-cell contains $9\times 9$ pillars (minus $2$ for the vacancies) with radii $\sigma=1/5$.
        	(c) Center frequency and (d) frequency difference vs pillar radius $\sigma$. Inset shows the linear coupling strength vs pillar radius. Circles denote values obtained from
        	the two lowest frequencies of the dimer, while plusses are estimates calculated from equation \eqref{eq:freq_estimate}
	in the tight-binding approximation.
        	The dashed line in (c) corresponds to frequency $\Omega = 2/(1-1.03\sigma)$, measured in units of $a^{-1}\sqrt{T_0/\rho}$ where $a$ is the lattice spacing, $\rho$ is the mass density of the membrane and $T_0$ is the pre-tension in the membrane. (e) Duffing constant, (f) quadratic and cubic coupling strength vs pillar radius $\sigma$.
        The dashed line in (e) shows Duffing constant $D_n = 4.5/(1-1.03\sigma)^6$, in units of $T_0/a^2$.}
	\label{fig:defmol_freqs}
	\label{fig:defmol_modes}
\end{figure*}

%%%%%%%%%%%%%%%%%%%%%%%%%%%%%%%%%%%%%%%%%%%%%%%%%%%%%%%%%%%%%%%%%%%%%%%%%%%%%%%
%% DEFECTS: PHONOMS AND PHONOCULES
%%%%%%%%%%%%%%%%%%%%%%%%%%%%%%%%%%%%%%%%%%%%%%%%%%%%%%%%%%%%%%%%%%%%%%%%%%%%%%%
\noindent
{\bf Coupled localized modes.} 
Considering two vacancies in the lattice, which are far apart, one finds two degenerate, localized modes.
Moving the vacancies closer together, the degeneracy is lifted and
one obtains a symmetric and an antisymmetric mode at frequencies $\Omega_\mp$.
Figures \ref{fig:defmol_freqs}(a) and \ref{fig:defmol_freqs}(b) show the mode shapes for $\sigma=1/5$. The frequency
splitting indicates that the localized modes leak into the region of the neighboring vacancy, which leads to an intermode interaction. 

Figures \ref{fig:defmol_freqs}(c) and \ref{fig:defmol_freqs}(d) show the center frequency $(\Omega_+ + \Omega_-)/2$ and the frequency difference $\Omega_+ - \Omega_-$ of a dimer for different pillar radii. The center frequency increases with increasing $\sigma$, since the the size of the suspended region shrinks. The frequency difference on the other hand decreases, because the overlap of the localized modes vanishes. For an estimate of the scaling behavior, one can consider the suspended region to be approximately a square of
size $\xi\times\xi$. From a dimensional analysis one finds, that the frequency scales as $\Omega \propto 1/\xi$. The length $\xi$ can be
taken as a linear function of $\sigma$, as shown in Fig.\ \ref{fig:defmol_freqs}(c).

The interaction strength is calculated from the linearized equation of motion obtained from equation \eqref{eq:Lagrangian}
by making the Ansatz $w(\mathbf{r}) = \psi_{\rm L}(\mathbf{r}) q_{\rm L} + \psi_{\rm R}(\mathbf{r}) q_{\rm R}$.
The localized mode shapes $\psi_n(\mathbf{r})$ are expressed by linear combinations of the exact dimer eigenmodes (Wannier basis)
or estimated by using mode shapes of the single-vacancy modes (tight-binding approximation)\cite{asme76}.
In both cases one obtains the generalized eigenvalue problem
\begin{equation}\label{eq:freq_estimate}
    \Omega_\pm^2 \sum_m S_{mn} q_m = \sum_m V_{nm} q_m\;\; (m,n)=({\rm L, R})\;,
\end{equation}
where the overlap matrix $S_{nm}=\braket{\psi_n}{\psi_m}$, the interaction matrix $V_{nm}=\braket{\nabla \psi_n}{\nabla \psi_m} + \braket{\psi_n}{V \psi_m}$ and angular brackets denote spatial integration. The diagonal terms $\Omega_n^2=V_{nn}$ give the individual mode-frequencies and the off-diagonal elements $\lambda_{nm}=V_{nm}$ are the linear coupling constants. In the tight-binding approximation, the overlap matrix $S_{nm}$ is not diagonal, but the values on the diagonal dominate. The effective interaction matrix is then obtained by multiplying $V_{nm}$ by $S_{nm}^{-1}$ from the left. From Figs. \ref{fig:defmol_freqs}(c) and \ref{fig:defmol_freqs}(d)
we find a good agreement between the tight-binding and Wannier calculations for the chosen values of $\sigma$. This implies that an array of such defect modes support frequency bands which are well described by the tight-binding method (see Supplementary Note 1).\\

%%%%%%%%%%%%%%%%%%%%%%%%%%%%%%%%%%%%%%%%%%%%%%%%%%%%%%%%%%%%%%%%%%%%%%%%%%%%%%%
%% COUPLED NONLINEAR OSCILLATORS
%%%%%%%%%%%%%%%%%%%%%%%%%%%%%%%%%%%%%%%%%%%%%%%%%%%%%%%%%%%%%%%%%%%%%%%%%%%%%%%
\noindent
{\bf Nonlinear effects.} One consequence of the membrane being atomically thin is that geometrically induced nonlinear effects become appreciable already for moderate displacements.
The respective contribution to the stretching energy in equation \eqref{eq:Lagrangian} is
\begin{equation}\label{eq:energy_nlin}
    E_{\rm nlin} = \frac{1}{8\varepsilon} \left\langle \left[ \left( \partial_x w \right)^2 + \left( \partial_y w \right)^2 \right]^2 \right\rangle \;.
\end{equation}
This expression accounts for the energy required to accommodate for the area increase associated with a transverse deflection of the membrane.

To estimate the influence of the energy \eqref{eq:energy_nlin} for the dimer, we use the Wannier basis, i.e., linear combinations of the eigenmodes for $w$. The energy is written as
\begin{multline}\label{eq:energy_nlin_amp}
    E_{\rm nlin} = \frac{1}{8\varepsilon}  \left( D_{\rm L} q_{\rm L}^4 + D_{\rm R} q_{\rm R}^4 + Q_{\rm L R} q_{\rm L}^2 q_{\rm R}^2 \right. \\
     \left.+C_{\rm L R} q_{\rm L} q_{\rm R}^3 + C_{\rm R L} q_{\rm R} q_{\rm L}^3 \right)\;,
\end{multline}
where $D_{\rm L, R}$, $C_{\rm L R}$ and $Q_{\rm L R}$ are the Duffing, the cubic coupling and the the quadratic coupling constants, respectively (see Methods section).
%%%%%%%%%%%%%%%%%%%%%%%%%%%%%%%%%%%%%%%%%%%%%%%%%%%%%%%%%%%%%%%%%%%%%%%%%%%%%%%%%%%%%

%%%%%%%%%%%%%%%%%%%%%%%%%%%%%%%%%%%%%%%%%%%%%%%%%%%%%%%%%%%%%%%%%%%%%%%%%%%%%%%%%%%%%
Figures \ref{fig:defmol_freqs}(e) and \ref{fig:defmol_freqs}(f) show the behavior of $D_{\rm L,R}$, $C_{\rm LR}$ and $Q_{\rm LR}$ as functions of $\sigma$. The Duffing constant increases with increasing $\sigma$, due to the shrinking suspension area. The nonlinear coupling constants, $C_{\rm LR}$ and $Q_{\rm LR}$, decrease analogous to the frequency splitting because the overlap of the localized modes vanishes. Using the same square-membrane approximation as for the center frequency, one finds the scaling $D_{\rm L, R} \propto 1/\xi^6$, which is in good agreement with the
numerical data.\\

%%%%%%%%%%%%%%%%%%%%%%%%%%%%%%%%%%%%%%%%%%%%%%%%%%%%%%%%%%%%%%%%%%%%%%%%%%%%%%%
%% RWA
%%%%%%%%%%%%%%%%%%%%%%%%%%%%%%%%%%%%%%%%%%%%%%%%%%%%%%%%%%%%%%%%%%%%%%%%%%%%%%%
\noindent
{\bf Dynamics in rotating wave approximation.} Since the time scales associated with $\Omega_\pm$ and the nonlinearities
are well separated, it is useful to consider the dynamics of the coupled oscillator system in the rotating wave approximation (RWA).
To this end we define $q_m = \sqrt{1 /2 \Omega_0} \left( a_m e^{-\imath \Omega_0 t} + a_m^* e^{\imath \Omega_0 t}\right)$
and $\dot{q}_m = -\imath \sqrt{{\Omega_0}/{2}} \left( a_m e^{-\imath \Omega_0 t} - a_m^* e^{\imath \Omega_0 t}\right)$,
where $\Omega_0$ is a reference frequency. The slowly changing amplitudes $a_m$ obey the equations of motion
$\dot{a}_m= -\imath {\partial H/\partial a_m^*}$ (see also Supplementary Note 2) \cite{brei12}. For a symmetric system, i.e., $D=D_{\rm L/R}$, $C=C_{\rm LR/RL}$, $Q=Q_{\rm LR/RL}$ and $\lambda=\lambda_{\rm LR/RL}$, the Hamiltonian $H$ in RWA is given by  
\begin{multline}\label{eq:H_RWA}
   H \approx  \frac{\delta \Omega_{\rm L}}{2} |a_{\rm L}|^2 + \frac{\delta \Omega_{\rm R}}{2} |a_{\rm R}|^2
   	+  \frac{U}{2}  \left(|a_{\rm L}|^4 + |a_{\rm R}|^4\right)   \\
      + \left[ J + 3W \left(|a_{\rm L}|^2 + |a_{\rm R}|^2\right)\right] \left[ a^*_{\rm L} a_{\rm R} + a^*_{\rm R} a_{\rm L} \right]\\
      + K\left[ |a_{\rm R}|^2 |a_{\rm L}|^2 + \frac{1}{4} \left( a_{\rm L}^2 (a_{\rm R}^*)^2 + a_{\rm R}^2 (a_{\rm L}^*)^2 \right)\right] \;,
\end{multline}
where $\delta \Omega_n = (\Omega_n^2-\Omega_0^2)/\Omega_0$ is a small detuning from the reference frequency, and we have defined $J=\lambda/\Omega_0$, $U=(3 D)/(8\varepsilon \Omega_0^2)$, $K=Q/(8\varepsilon \Omega_0^2)$ and
$W=(3 C)/(32\varepsilon \Omega_0^2)$ to characterize ``hopping'', on-site interactions and interaction
tunneling, respectively. The values of $J$, $U$, $K$ and $W$ following from Fig.\ \ref{fig:defmol_freqs} are shown in Supplementary Fig.\ 2.

Equation \eqref{eq:H_RWA} corresponds to the classical analog of the ``improved two-mode model'' \cite{anbe06} for Bose particles in a double-well potential. The Hamiltonian \eqref{eq:H_RWA} can be expressed in terms of only two variables. 
Setting $a_n = |a_n| e^{i\phi_n}$, we define the ``population imbalance'' $z=(|a_{\rm R}|^2 - |a_{\rm L}|^2)/Z^2$ and the relative phase $\Delta \phi=\phi_{\rm R}-\phi_{\rm L}$. Note that the ``total population'' $Z^2=|a_{\rm R}|^2 + |a_{\rm L}|^2$ is conserved \cite{eijo03}. We further set the reference frequency to the average frequency, $\Omega_0^2 = (\Omega_{\rm L}^2+\Omega_{\rm R}^2)/2$, which maps equation \eqref{eq:H_RWA} onto the Hamiltonian for a rigid rotor\cite{smfa+97,anbe06}
\begin{multline}\label{eq:eff_H}
   H_{\rm eff} = \frac{U_{\rm eff} z^2}{2} + J_{\rm eff}\sqrt{1-z^2} \cos(\Delta \phi)  \\
       + \frac{K_{\rm eff}}{2}(1-z^2)\cos(2\Delta \phi) - \mu_{\rm eff} z\;,
\end{multline}
where $U_{\rm eff}=(U-K)Z^4/2$, $J_{\rm eff}=(J + W Z^2) Z^2$, $K_{\rm eff}=K Z^4/4$ and $\mu_{\rm eff}=(\Omega_{\rm L}^2 -\Omega_{\rm R}^2)Z^2/4\Omega_0$. The last term is due to detuning of the individual mode frequencies and tilts the double-well potential defined by this Hamiltonian. In the present system, $K_{\rm eff}\ll {\rm max}(U_{\rm eff}, J_{\rm eff})$ (see Supplementary Fig.\ 2), and the second to last term in equation \eqref{eq:eff_H} does not crucially influence the dynamics.
Taking $K_{\rm eff}\to0$, the corresponding equations of motion are
\begin{subequations}
\begin{align}\label{eq:eff_eom}
   \dot{z} ={}&  J_{\rm eff}\sqrt{1-z^2} \sin(\Delta \phi)\;, \\
   \Delta \dot{\phi} ={}& U_{\rm eff} z - J_{\rm eff} \frac{z}{\sqrt{1-z^2}}\cos(\Delta \phi) - \mu_{\rm eff}\;.
\end{align}
\end{subequations}

For atoms in optical lattices, equation \eqref{eq:eff_H} is usually obtained in the semiclassical limit, for example by using the Gutzwiller Ansatz. Here, we derived the Hamiltonian \eqref{eq:eff_H} within a completely classical description employing the RWA (see also Supplementary Note 2).
As a consequence, the total population $Z$ used above has the unit of an action and is proportional, but not equal to the total number of phonons.

\begin{figure*}[ht]
    \centering
    \includegraphics[width=0.9\textwidth]{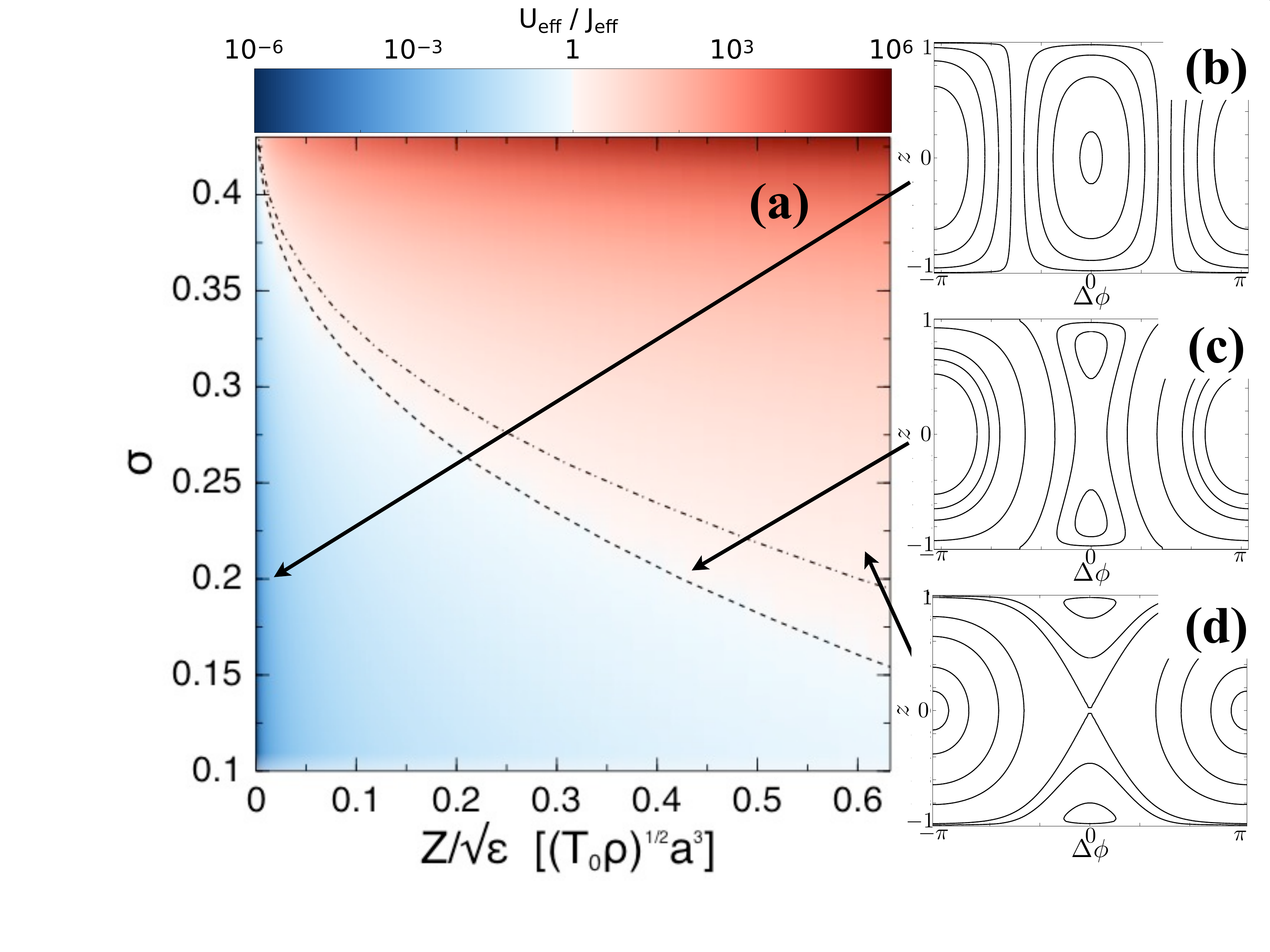}
\caption{Dynamical regimes of equation \eqref{eq:eff_H}: (a) Ratio of nonlinear to linear interaction $U_{\rm eff}/J_{\rm eff}$ as a function of ``total population'' $Z$ and pillar radius $\sigma$. By tuning the vibrational amplitude of the system, both the ``Rabi'' (blue) and ``Josephson'' (red) regimes are accessible The dashed line corresponds to $U_{\rm eff}/J_{\rm eff}=1$, while the dashed-dotted line corresponds to $U_{\rm eff}/J_{\rm eff}=2$. (b-d) Phase space plots of corresponding to the different regimes for tilt $\mu_{\rm eff}=0$. (b) At low values of the ratio ($U_{\rm eff}/J_{\rm eff}= 1/10$), the system undergoes Rabi-like oscillations with vanishing temporal mean of the population imbalance $z$. $\left(c\right)$: For values of the ratio $1<U_{\rm eff}/J_{\rm eff}<2$ (here $U_{\rm eff}/J_{\rm eff}= 3/2$), the system develops two elliptic fixed points with $z$ having non vanishing temporal mean. (d) For higher ratios (here $U_{\rm eff}/J_{\rm eff}= 3$) running phase modes appear.}
\label{fig:BHDRegimes}
\end{figure*}

For $\mu_{\rm eff}=0$, the system undergoes a pitch-fork bifurcation at a critical value of the ratio $U_{\rm eff}/J_{\rm eff}$, with qualitatively different dynamics on each side of the bifurcation. This defines two regimes, denoted in the literature as the ``Rabi'' regime for $U_{\rm eff}/J_{\rm eff}\ll 1$ and ``Josephson'' regime for $U_{\rm eff}/J_{\rm eff}\gg 1$ \cite{zini+10}. The Rabi regime is characterized by oscillation of $z$ with vanishing temporal mean \cite{fari+13}. In the Josephson regime there exist self-trapped fixed points at $z=\pm z_0$ and $\Delta \phi=0$, corresponding to either of the localized modes being predominantly excited. At even larger ratios ($U_{\rm eff}/J_{\rm eff}>2$), running phase modes appear (Figs. \ref{fig:BHDRegimes}(b-d)).  

Note the different dependence of $U_{\rm eff}$ and $J_{\rm eff}$ on the vibrational amplitudes, entering through the parameter $Z$. This implies that the ratio can be tuned either through changes in the system itself (through the pillar radii), or by changing the vibrational amplitude of the dimer. In Fig. \ref{fig:BHDRegimes}(a) the ratio $U_{\rm eff}/J_{\rm eff}$ is shown as a function of the pillar radius and the vibrational amplitude scaled by the pre strain. The dashed line corresponds to the transition between the two regimes. At a fixed pillar size, the system can be tuned into either regime through the vibrational amplitude.\\

%%%%%%%%%%%%%%%%%%%%%%%%%%%%%%%%%%%%%%%%%%%%%%%%%%%%%%%%%%%%%%%%%%%%%%%%%%%%%%%
%% TUNABILITY
%%%%%%%%%%%%%%%%%%%%%%%%%%%%%%%%%%%%%%%%%%%%%%%%%%%%%%%%%%%%%%%%%%%%%%%%%%%%%%%
\noindent
{\bf Frequency tuning.}
External forces applied to the membrane will lead to a static deformation of the membrane. Due to the nonlinearities of the system, this can be used to tune the frequencies and couplings. Considering small oscillations around the static deformation, the equations of motion are linearized, which
leads to the following renormalized frequencies and linear couplings
\begin{subequations}\label{eq:tuning}
\begin{align}
    {\Omega}^2_n &{}\to {\Omega}_n^2 + \frac{3}{2\varepsilon} \bar{q}_n^2 D_n + \frac{1}{4\varepsilon} Q_{n m}\bar{q}_m^2 + \frac{3}{4\varepsilon} C_{n m} \bar{q}_m \bar{q}_n \;,\\
    {\lambda}_{nm} &{}\to {\lambda}_{nm} + \frac{1}{2\varepsilon}  Q_{m n} \bar{q}_n\bar{q}_m + \frac{3}{8\varepsilon}  C_{n m} \left(\bar{q}_m^2 + \bar{q}_n^2\right)\;,
\end{align}
\end{subequations}
where $\bar{q}_{n}$ are the static deflections. Assuming that the latter can be tuned independently for each oscillator, for instance via local back gates, this allows for complete control over the frequencies and the splitting. This effect is interesting for applications and may be used to study Landau-Zener transitions\cite{fari+12} beyond the linear regime \cite{biqi00}.

\begin{figure}[ht]
    \centering
    \includegraphics[width=\columnwidth]{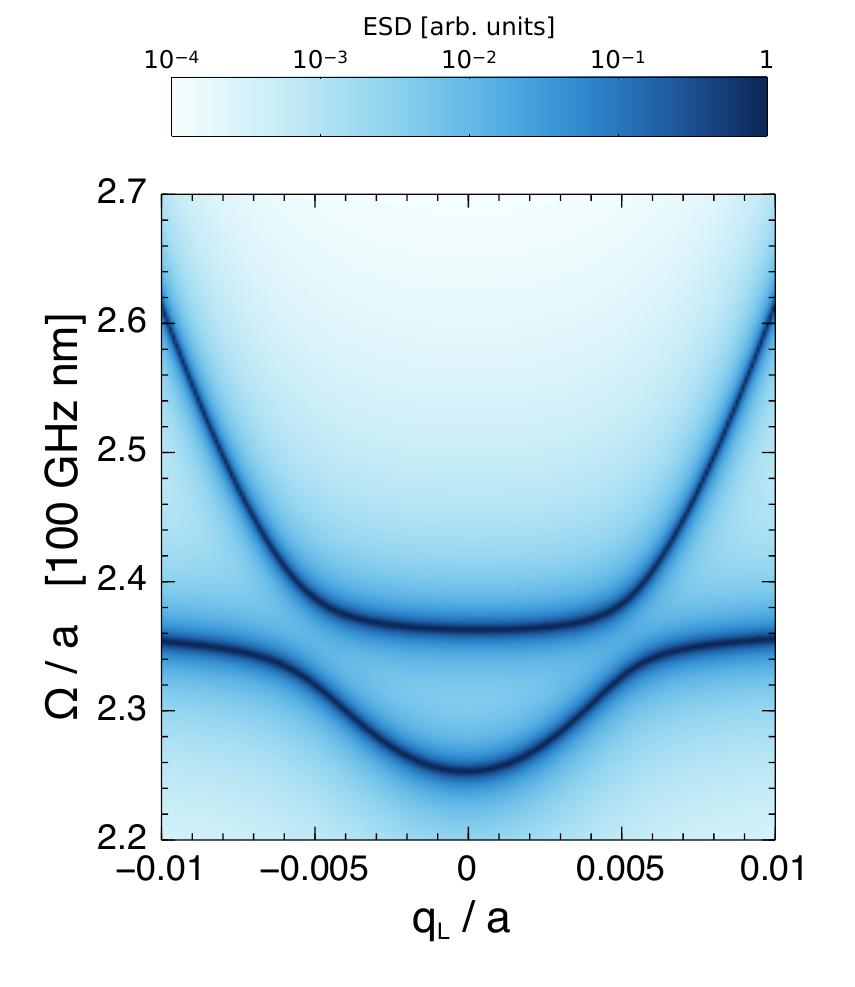}
\caption{Frequency tuning: Energy spectral density ($ESD(\Omega)$) for the dimer as a function of static displacement $q_{\rm L}$ of the left mode and the driving frequency $\Omega$ for fixed static displacement $q_{\rm R}=0.005 a$, where $a$ is the lattice spacing. The data is scaled such that ${\rm max}[ESD(\Omega)]=1$. Parameters are: pillar radius $\sigma=1/5$, quality factor $Q\approx500$ and pre-strain $\varepsilon=10^{-3}$.}
\label{fig:ESD}
\end{figure}

To demonstrate the frequency tuning, we consider a harmonically driven dimer. The energy spectral density per unit driving amplitude $ESD(\Omega)=\sum_n \Omega^2\,  |\chi_n(\Omega)|^2$, where $\chi_n(\Omega)$ is the susceptibility for mode $n$, gives a measure of the response at the driving frequency $\Omega$.
In Fig.\ \ref{fig:ESD}, the energy spectral density is shown as a function of the back-gate induced static deflection. 
Here, $\bar{q}_{\rm R}$ is kept constant while $\bar{q}_{\rm L}$ is swept. Further, we assumed a quality factor of $Q\approx500$
to account for damping of the mechanical motion.

%%%%%%%%%%%%%%%%%%%%%%%%%%%%%%%%%%%%%%%%%%%%%%%%%%%%%%%%%%%%%%%%%%%%%%%%%%%%%%%
%% SUMMARY
%%%%%%%%%%%%%%%%%%%%%%%%%%%%%%%%%%%%%%%%%%%%%%%%%%%%%%%%%%%%%%%%%%%%%%%%%%%%%%%
\section{Discussion}
We have investigated the realization of a new type of phononic crystal, using periodically pinned
and atomically thin membranes. Distorting the periodicity of the pinning lattice, allows it to support coupled localized modes.
Due to the geometric nonlinearity the dynamics of the mode-amplitudes is described by coupled
nonlinear oscillator equations. By applying external forces, the nonlinearities allow tuning the frequencies and couplings.

Using typical values for graphene resonators\cite{lewe+08,peue10,kusc+01,limi+12}, $\lambda+2\mu\approx340\;{\rm N m^{-1}}$, $\kappa\approx1.5\;{\rm eV}$, 
$\rho\approx 7.6\cdot 10^{-7}{\rm kg m^{-2}}$, $V_0\approx 1.8\cdot10^{20}\;{\rm N m^{-3}}$ and $\varepsilon=0.1\%$ one finds for the unit of frequency $100/a\;{\rm GHz\; nm}$ and the dimensionless pinning strength $a^2 V_0/T_0=500 a^2\; {\rm nm^{-2}}$. Assuming a lattice spacing of $100\;{\rm nm}$ therefore leads to frequencies in the GHz range.
The graphene sheet can be treated as a membrane, if the bending energy in equation \eqref{eq:Lagrangian} can be neglected.
In our case, this contribution becomes negligible for $a\gg \sqrt{\kappa/\varepsilon(\lambda+2\mu)}$, which corresponds to $a\gg8{\rm \AA}$. Then, the only parameters entering the problem are the details of the pinning potential, more specifically the pinning strength and the pillar radius. In the limit of strong pinning, the membrane will essentially be clamped at the pillars ($w=0$). Recently, structures with graphene sheets suspended between silica pillars have been experimentally realized \cite{pldi+14}.

A natural extension of the dimer configuration considered in the Results is a periodic array of defect modes. Analogous to the dimer, the energy of such a system can be expressed in terms of the amplitudes of localized (Wannier) modes.
In general, the resulting Hamiltonian in RWA is given by
\begin{multline}\label{eq:H_latt}
	H = \sum_n \Bigl( \frac{\delta \Omega_n}{2} |a_n|^2
       +\sum_{m\neq n} \frac{J_{m n}}{2} \left[ a^*_m a_n + a^*_n a_m \right] \\
       +\sum_{m,i,j} W_{nm,ij} a^*_n a^*_m a_i a_j \Bigr)\;,
\end{multline}
where $J_{nm}$ and $W_{nm,ij}$ denote ``hopping'' and interaction constants, respectively.
For a square lattice with dominating nearest neighbor couplings the interaction energy can be constructed from the dimer contributions (see Supplementary Note 1). In that case, the hopping is restricted to nearest neighbors and the interaction term only contains contributions from at most two neighboring modes. More general types of interactions can be realized by using other lattice configurations.
The equations of motion for the mode amplitudes ensuing from equation \eqref{eq:H_latt} are the classical version of a discrete nonlinear Schr\"{o}dinger equation, known to display rich nonlinear phenomena \cite{eijo03,eilo+85}.

To actuate the motion of the oscillators, external pressure may be applied to the membrane. In NEMS this is typically achieved by
the electro-static interaction with a metallic back-gate\cite{sook+12}. This leads to the pressure $P_z\propto -G(\mathbf{r}) V_{\rm g}^2$ with $V_{\rm g}$ denoting the gate voltage and $G(\mathbf{r})$ the shape of the gate\cite{atis+08}. The force acting on each oscillator is given by $f^{(\rm ext)}_n \propto -\braket{\psi_n}{G} V_{\rm g}^2$. A DC voltage leads to a static displacement
$\bar{q}_{n}>0$ in equations \eqref{eq:tuning}, while an AC voltage can be used to drive the system.

The coupling of the membrane to its environment leads to dissipation of the mechanical motion. This has been extensively studied in the context of atomically thin membrane resonators\cite{immo14}, and quality factors of $Q>100,000$ have been reported for graphene resonators at low temperatures \cite{eimo+11,wegu+14}. Phenomenologically, the dissipation can be described by a viscous damping term proportional to $Q^{-1} \dot{q}_n$ in the equations of motion. In the Rabi regime the oscillation frequency is given by $|J|$
and $Q\gg1/|J|$ is required. We find for $\sigma=1/5$, that $1/|J|\approx 65$ and $Q\approx500$ would be sufficient to resolve the oscillations. In addition to the viscous damping, nonlinear dissipation has been observed in carbon based resonators\cite{eimo+11}, which might give rise to interesting nonlinear dissipation effects\cite{licr08}.

Due to the high frequencies of the localized modes and the low mass of the membrane, our system is a promising candidate for
studies in the quantum regime\cite{poza12}. The presence of intrinsic nonlinearities makes it attractive for creation and detection
of non-classical states\cite*{voki+12,vocr+13,vois+13}. More generally,
realizing lattices of interacting flexural modes would pave the road for {\em quantum many-body phononics}.

%%%%%%%%%%%%%%%%%%%%%%%%%%%%%%%%%%%%%%%%%%%%%%%%%%%%%%%%%%%%%%%%%%%%%%%%%%%%%%%
%% METHODS
%%%%%%%%%%%%%%%%%%%%%%%%%%%%%%%%%%%%%%%%%%%%%%%%%%%%%%%%%%%%%%%%%%%%%%%%%%%%%%%
\section{Methods}
\noindent
{\bf Nonlinear constants.} Using the Ansatz $w(\mathbf{r}) = \psi_{\rm L}(\mathbf{r}) q_{\rm L} + \psi_{\rm R}(\mathbf{r}) q_{\rm R}$
in equation \eqref{eq:energy_nlin} leads to the expression of the energy \eqref{eq:energy_nlin_amp} in terms
of the nonlinear constants
\begin{subequations}
\begin{align}
    D_n ={}& \left\langle \left[ \left( \partial_x \psi_n \right)^2 + \left( \partial_y \psi_n \right)^2 \right]^2 \right\rangle\;,\\
    C_{\rm RL} ={}& 4\left\langle
                          \left[
                                  \left( \partial_x \psi_{\rm L} \right)^2 + \left( \partial_y \psi_{\rm L} \right)^2
                          \right] \right.\notag\\
                   {}&\times
                          \left[
                              \left( \partial_x \psi_{\rm L} \right)\left( \partial_x \psi_{\rm R} \right) + \left( \partial_y \psi_{\rm L} \right)\left( \partial_y \psi_{\rm R} \right) \right]
                         \Big \rangle\;,\\
    C_{\rm LR} ={}& 4\left\langle
                          \left[
                                  \left( \partial_x \psi_{\rm R} \right)^2 + \left( \partial_y \psi_{\rm R} \right)^2
                          \right]\right.\notag\\
                   {}&\times
                          \left[
                              \left( \partial_x \psi_{\rm R} \right)\left( \partial_x \psi_{\rm L} \right)
                              + \left( \partial_y \psi_{\rm R} \right)\left( \partial_y \psi_{\rm L} \right) \right]
                         \Big \rangle\;,\\
    Q_{\rm LR} ={}& 2\Big \langle 4\left( \partial_x \psi_{\rm L} \right)\left( \partial_x \psi_{\rm R} \right)\left( \partial_y \psi_{\rm L} \right)\left( \partial_y \psi_{\rm R} \right) \notag\\
                    {}&
                            + \left( \partial_x \psi_{\rm L} \right)^2
                              \left[
                                      \left(\partial_y \psi_{\rm R} \right)^2 + 3 \left(\partial_x \psi_{\rm R} \right)^2
                              \right] \notag\\
                    {}&    \left.
                            + \left( \partial_y \psi_{\rm L} \right)^2
                              \left[
                                      \left(\partial_x \psi_{\rm R} \right)^2 + 3 \left(\partial_y \psi_{\rm R} \right)^2
                              \right]
                          \right\rangle\;.
\end{align}
\end{subequations}

\section{Acknowledgements}\noindent
The authors acknowledge funding from the Swedish Research Council and from the European
Union through project no.\ 604391 (Graphene Flagship).

%%%%%%%%%%%%%%%%%%%%%%%%%%%%%%%%%%%%%%%%%%%%%%%%%%%%%%%%%%%%%%%%%%%%%%%%%%%%%%%
%% REFERENCES
%%%%%%%%%%%%%%%%%%%%%%%%%%%%%%%%%%%%%%%%%%%%%%%%%%%%%%%%%%%%%%%%%%%%%%%%%%%%%%%
%\bibliographystyle{naturemag}
%\bibliography{grphononics}

%\end{document}
%\cleardoublepage
%\begin{document}
\pagebreak
\begin{widetext}
{\large \bf Supplementary material for nonlinear phononics using atomically thin membranes}
%\title{Supplement for nonlinear phononics using atomically thin membranes}
%\author{Daniel Midtvedt}
%\email{midtvedt@chalmers.se}
%\author{Andreas Isacsson}
%\author{Alexander Croy}
%\email{croy@chalmers.se}
%
%\affiliation{Department of Applied Physics,
%Chalmers University of Technology, S-41296 G\"{o}teborg, Sweden}
%
%\date{\today}

%\pacs{%
%}
%
%\begin{abstract}\noindent
%\end{abstract}
%
%\maketitle
%%
\begin{figure*}[ht]
	\centering
	\begin{minipage}[c]{0.5\textwidth}
		\includegraphics[width=0.95\textwidth]{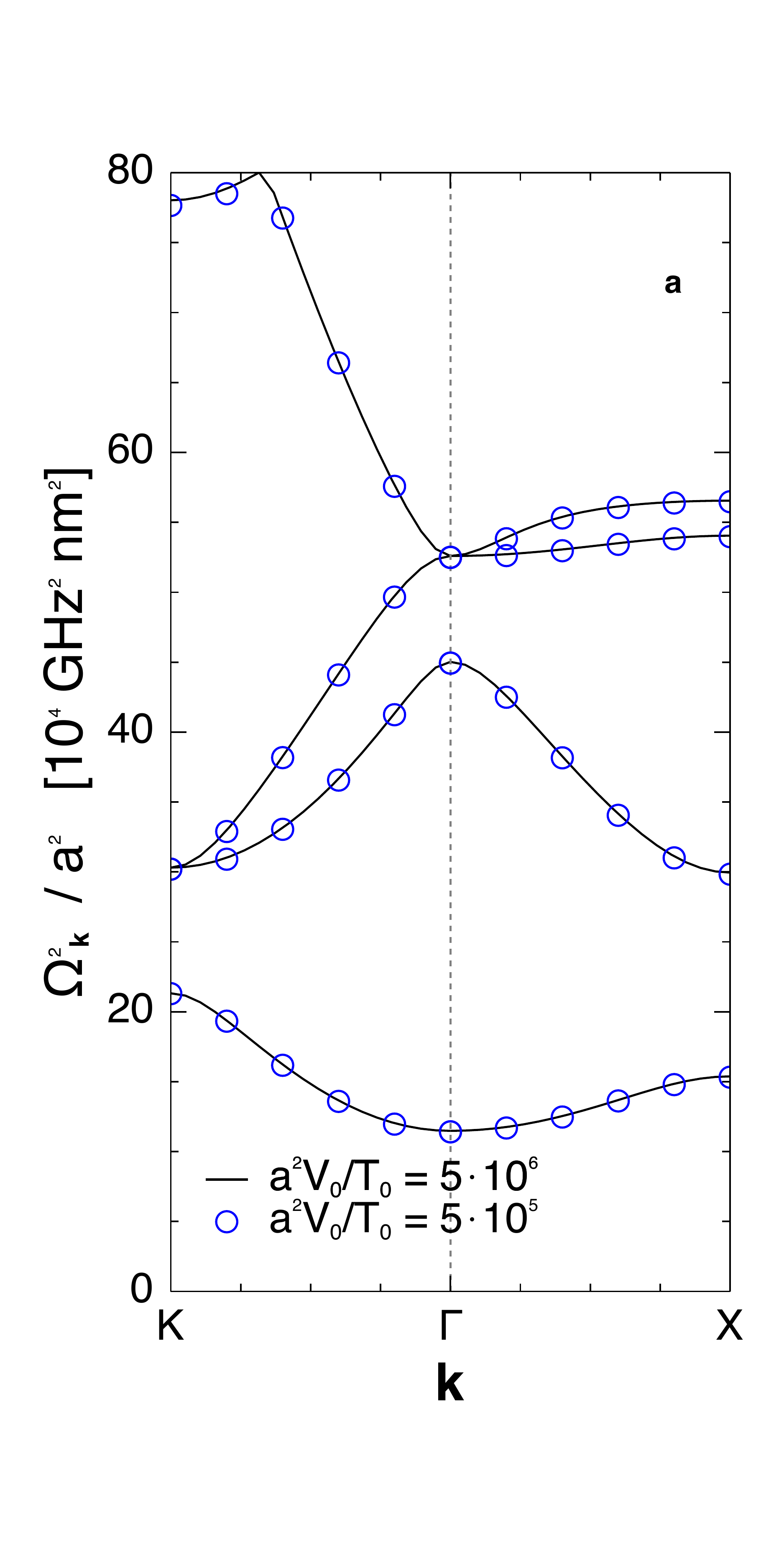}
	\end{minipage}\hfill
	\begin{minipage}[c]{0.4\textwidth}
		\centering
		\includegraphics[width=0.6\textwidth]{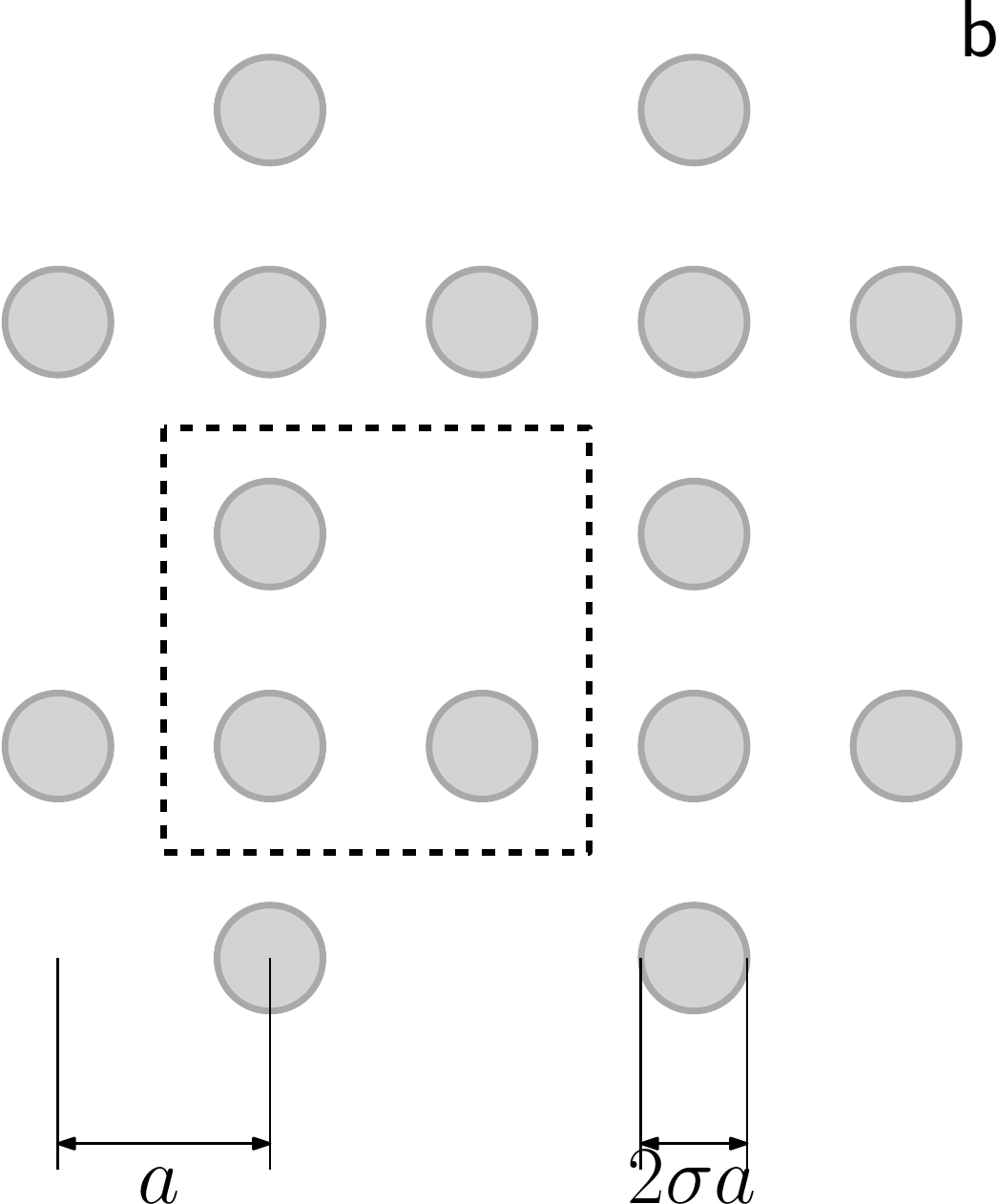}\\\vspace{5ex}
		\includegraphics[width=\textwidth]{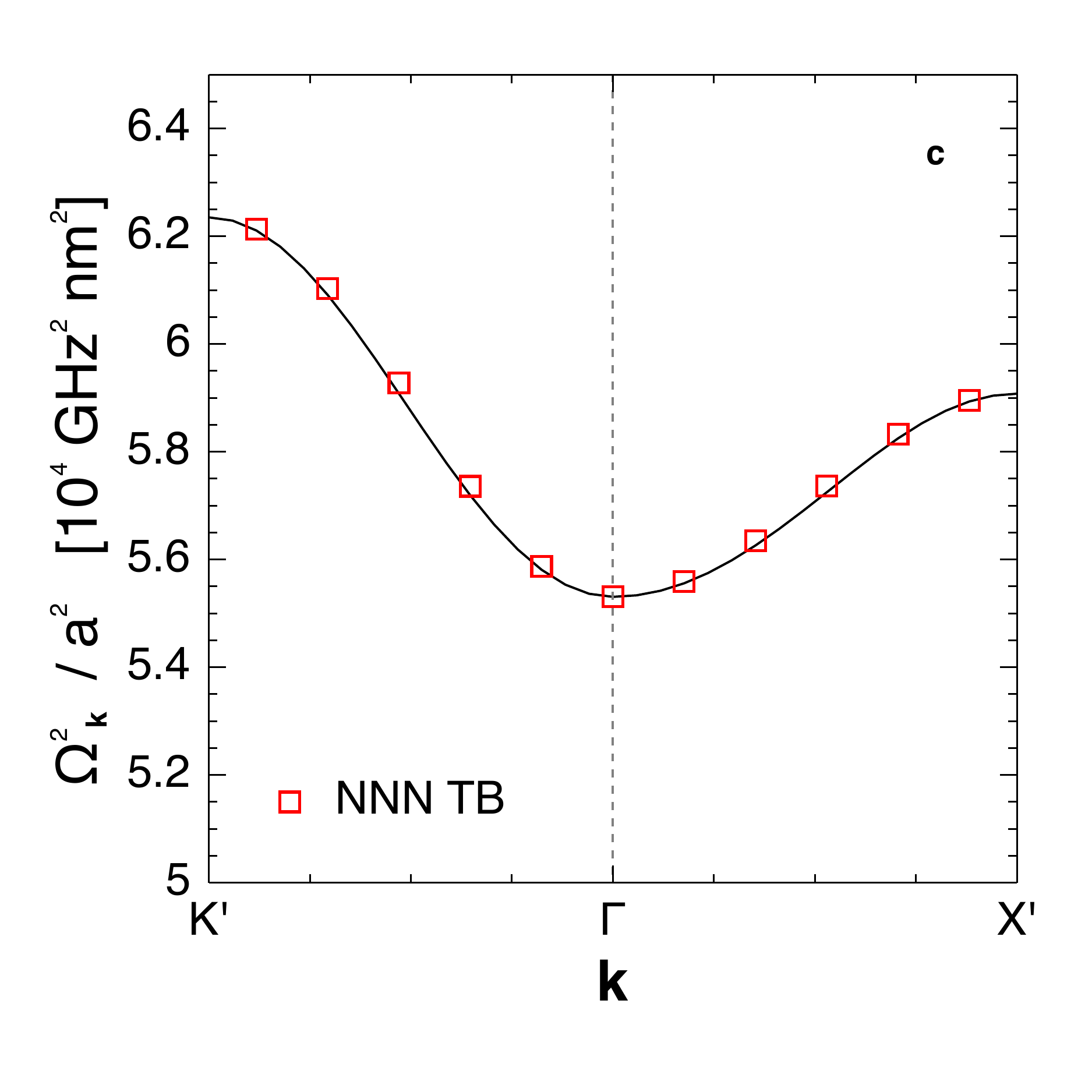}
	\end{minipage}
	\caption{Frequency bands for the periodically pinned membrane: (a) The pillars have radii $\sigma=1/5$ and pinning strengths $a^2 V_0/T_0=5\cdot10^5$ (circles) and $a^2 V_0/T_0=5\cdot10^6$ (full lines), where $a$ is the lattice spacing and $T_0$ is the pre-tension of the membrane. (b) Schematic of a square lattice of single-vacancy defects. The unit cell is highlighted by the dashed rectangle. The periodicity of the lattice is $2a$. (c) The lowest frequency band of the defect lattice. This band is situated well below the lowest band of the defect-free lattice. Symbols denote bands of a tight-binding model with
	next-nearest neighbor coupling and parameters $\lambda=0.09 a^{-2} T_0/\rho$, $\lambda'=0.003 a^{-2} T_0/\rho$ and $\Omega_{\rm c}^2=5.9 a^{-2} T_0/\rho$, where $\rho$ is the mass density of the membrane, which were extracted from the full band calculation.}
	\label{fig:bands}
	\label{fig:deflat}
\end{figure*}
\begin{figure*}[ht]
	\centering
		\includegraphics[width=0.95\textwidth]{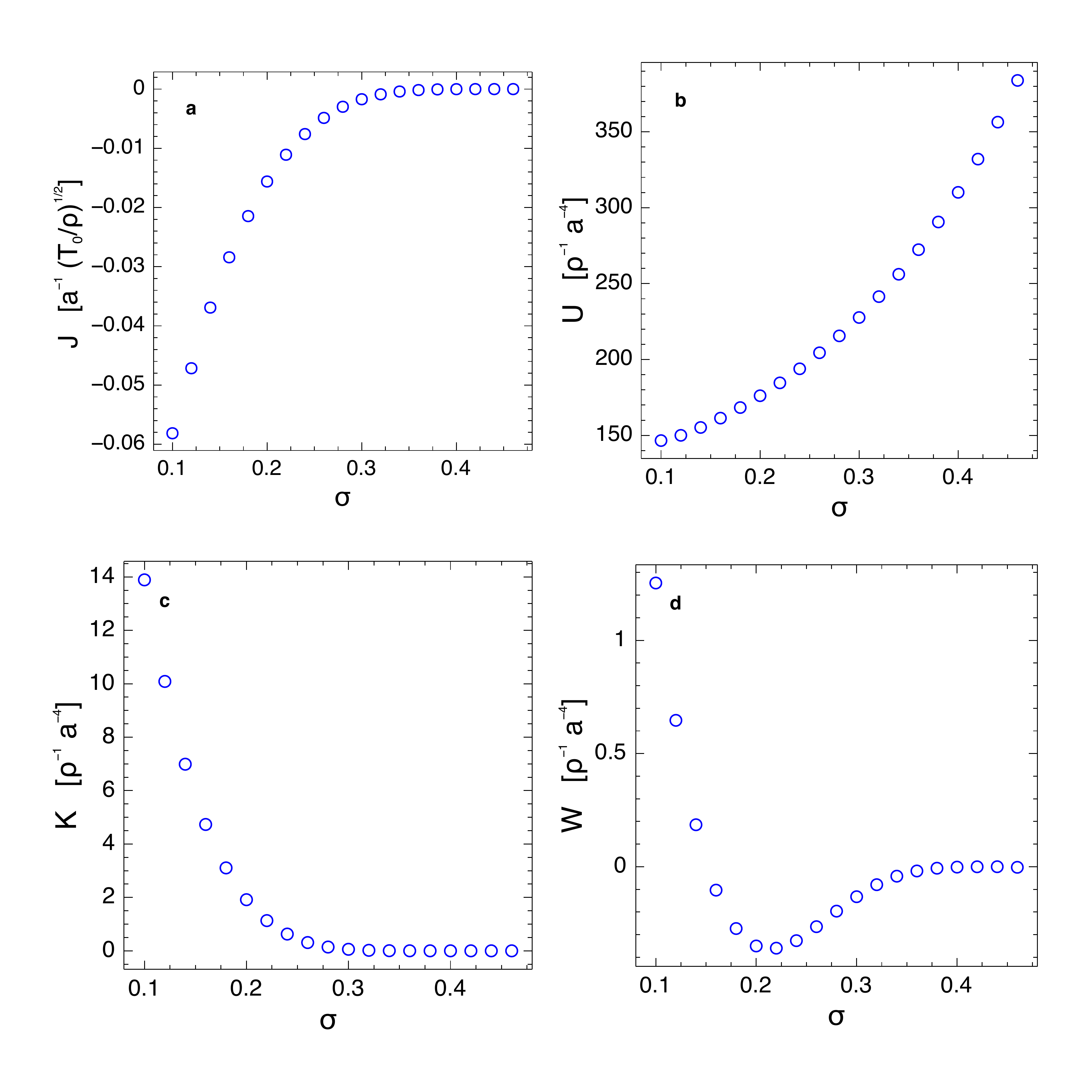}
	\caption{Dimer parameters entering the RWA Hamiltonian: (a) hopping parameter $J=\lambda/\Omega_0$, (b) and (c) onsite interactions
		$U=3D/8\varepsilon\Omega_0^2$, $K=Q/8\varepsilon\Omega_0^2$
		and (d) interaction tunneling $W=3C/32\varepsilon\Omega_0^2$ for pre-strain $\varepsilon=10^{-3}$ as functions of pillar radius $\sigma$.
		The values of $\Omega_0$, $\lambda$, $D$, $C$ and $Q$ are given in Fig.\ 2.}
	\label{fig:dimer}
\end{figure*}

\clearpage
\section{Lattice of vacancies}
A lattice of vacancies support frequency bands, much like the electronic bands in periodic solids. To calculate the frequency bands $\Omega(\mathbf{k})$ we linearize the
equation of motion for $w$ resulting from the Lagrangian (1),
\begin{equation}
	\Omega^2 w(\mathbf{r}) = - \nabla^2 w(\mathbf{r}) + V(\mathbf{r}) w(\mathbf{r}) \;.
\end{equation}
Using the Bloch theorem yields the eigenvalue equation
\begin{equation}
	\label{eq:bloch}
	\Omega^2(\mathbf{k}) w(\mathbf{k}, \mathbf{r}) = \left[ -\left(\nabla + \imath \mathbf{k}\right)\cdot\left(\nabla + \imath \mathbf{k}\right) + V(\mathbf{r}) \right]w(\mathbf{k}, \mathbf{r})\;.
\end{equation}
To compute $\Omega^2(\mathbf{k})$ we use a finite-difference representation of $\nabla$, $V(\mathbf{r})$ and $w(\mathbf{k}, \mathbf{r})$ and diagonalize the resulting eigenvalue problem.\\

The first four frequency bands are shown in Supplementary Figure \ref{fig:bands}
for $\sigma=1/5$ and two different pinning strengths. The pinning potential breaks the
translational symmetry of the free graphene sheet, and thus a frequency
gap opens at the $\Gamma$-point. The frequency bands shift upwards with increasing potential strength $V_0$ until they saturate for
large $V_0$, since the membrane becomes fully clamped at the pillars.

As mentioned in the main text, the proposed system is a platform for creating lattices of localized defect modes by removing pillars in a periodic fashion. The resulting structure can again be described in terms of the Bloch theorem and frequency bands can be calculated as outlined above. As a simple example, we consider a square lattice constructed from the single vacancy defects introduced in the main text (see Supplementary Figure \ref{fig:deflat}(b)). Supplementary Figure \ref{fig:deflat}(c) shows the lowest frequency band for such a lattice with $\sigma=1/5$. 
This band is situated well below the lowest band of the defect-free pinning lattice. Note that in this case, the periodicity of the lattice is doubled, which shifts the X and K points to X' and K'. 

We compare the frequency band to the respective band obtained from a tight-binding model with
next-nearest neighbor couplings. In this case, one finds
\begin{multline}
	\Omega^2(k_x, k_y) = \Omega_{\rm c}^2 + 2\lambda\left[\cos\left(2 k_x a\right) + \cos\left(2 k_y a\right) \right]
													  + 2\lambda'\left[\cos\left(2 (k_x+k_y) a\right) + \cos\left(2 (k_x - k_y) a\right) \right]
\end{multline}
with $\lambda$ denoting the nearest-neighbor and $\lambda'$ the next-nearest neighbor coupling. From the full band calculation
we obtain $\lambda=0.09a^{-2} T_0/\rho$, $\lambda'=0.003a^{-2} T_0/\rho$ and $\Omega_{\rm c}^2=5.9a^{-2} T_0/\rho$, which is in very good agreement with the values
found for the dimer in the main text.

The small next-nearest neighbor coupling, $\lambda'\ll\lambda$, shows that the Wannier modes are well localized. In that case, 
the nonlinear contribution to the stretching energy given by equation (3) in the main text, approximately only contains contributions from at
most two neighboring modes. One finds
\begin{equation}\label{eq:energy_nlin_amp}
    E_{\rm nlin} \approx \frac{1}{8\varepsilon} \sum_n \left( D_n q_n^4 + \sum_{\mean{m\neq n}} C_{n m} q_m q_n^3 + \frac{1}{2} Q_{n m} q_m^2 q_n^2 \right)\;,
\end{equation}
where the sum over $m$ is restricted to nearest neighbors. Other terms only involve smaller next-nearest neighbor couplings. The constants $D_n$, $C_{nm}$ and $Q_{nm}$ are calculated according
to equation (9) in the main text. In the rotating wave approximation one obtains for the Hamiltonian
\begin{multline}\label{eq:H_RWA}
   H \approx \sum_n \left( \frac{\delta \Omega_n}{2} |a_n|^2 +
           \sum_{\mean{m\neq n}} \frac{J_{m n}}{2} \left[ a^*_m a_n + a^*_n a_m \right] 
           +  \frac{U_{n}}{2} |a_n|^4
           + \sum_{\mean{m\neq n}} \frac{K_{nm}}{2}  |a_m|^2 |a_n|^2
       \right) \\
       + \sum_n\sum_{\mean{m\neq n}} \left(
           \frac{K_{nm}}{8} \left[ a_m^2 (a_n^*)^2 + a_n^2 (a_m^*)^2 \right]
       +  3 W_{nm} |a_n|^2 \left[ a_m a_n^* + a_m^* a_n \right]
       \right) \;,
\end{multline}
where $\delta \Omega_n = (\Omega_n^2-\Omega_0^2)/\Omega_0$ is a small detuning from the reference frequency, and we have defined $J_{nm}=\lambda_{nm}/\Omega_0$, $U_{m}=(3 D_{m})/(8\varepsilon \Omega_0^2)$, $K_{nm}=Q_{nm}/(8\varepsilon \Omega_0^2)$ and
$W_{nm}=(3 C_{nm})/(32\varepsilon \Omega_0^2)$.

Supplementary Figure \ref{fig:dimer} shows the parameters $J$, $U$, $K$ and $W$ for $\varepsilon=10^{-3}$, which enter the RWA Hamiltonian (5) given in the main text. It can be seen, that $U\gg K$, which implies $U_{\rm eff}\gg K_{\rm eff}$.

\section{Rotating wave approximation}\label{sec:RWA}
The time evolution of an arbitrary function $f(p_n,q_n,t)$ is given by
\begin{equation}
	\frac{ d f}{dt}=\frac{\partial f}{\partial t} + \{f,E\},
\end{equation}
where $\{\cdot,\cdot\}$ is the Poisson bracket and $E(\{p_n\}, \{q_n\})$ is the Hamiltonian of the system. The equations of motion for the slowly varying amplitudes $a_n$ are derived by writing $\partial/\partial q_n=(\partial a_n/\partial q_n)\partial/\partial a_n + (\partial a^*_n/\partial q_n)\partial/\partial a^*_n$ and similarly for $\partial/\partial p_n$, which yields
\begin{equation}
	\frac{ d a_n}{dt}=\frac{\partial a_n}{\partial t} + \{a_n,E\}=\frac{\partial a_n}{\partial t} - \imath \frac{\partial E}{\partial a_n^*},
\end{equation}
due to the explicit time dependence in $a_n=\sqrt{\Omega_0/2T_0 a^2}\left(q_n+\imath p_n/\Omega_0\right) e^{\imath \Omega_0 t}$. By defining $H=E-\Omega_0 \sum_n |a_n|^2$ the equations of motion are more analogous to the ordinary Hamilton equations of motion,
\begin{equation}
	\dot{a}_n= -\imath \frac{\partial H}{\partial a_n^*}.
\end{equation}
The RWA Hamiltonian \eqref{eq:H_RWA} is obtained by neglecting oscillating terms in $H$.
\end{widetext}
\end{document}